\documentclass[letterpaper, 10 pt, conference]{IEEE/ieeeconf}  

\IEEEoverridecommandlockouts                              
\usepackage{amssymb}
\usepackage{amsmath}
\usepackage{bbm}
\usepackage{physics}
\usepackage{braket}
\usepackage{breqn}
\usepackage{tcolorbox}
\usepackage[hidelinks]{hyperref}
\usepackage{graphicx}
\usepackage{subcaption}
\usepackage{caption}
\let\labelindent\relax
\usepackage{enumitem}

\newcommand{\X}{\mathcal{X}}
\newcommand{\Xbf}{\mathbf{X}}
\newcommand{\xbf}{\mathbf{x}}
\renewcommand{\S}{\mathcal{S}}
\renewcommand{\O}{\mathcal{O}}
\newcommand{\ntraindelta}{m}
\newcommand{\ntestdelta}{N}
\newcommand{\ntraintime}{n}

\newcommand{\ncontrols}{c}
\newcommand{\dimZ}{z}
\newcommand{\dimX}{d}
\newcommand{\dimU}{p}
\newcommand{\nlcomb}{q}

\newcommand{\K}{\mathcal{L}}
\newcommand{\Kfin}{\mathcal{K}^{\Delta t}}
\newcommand{\KfinbfU}[1]{\mathbf{K}^{\Delta t}_{\left[ #1 \right]}}
\newcommand{\Kfinbf}{\mathbf{K}^{\Delta t}}
\newcommand{\ubf}{\mathbf{u}}
\newcommand{\U}{\mathcal{U}}
\newcommand{\zerobf}{\mathbf{0}}
\newcommand{\B}{\mathcal{B}}
\newcommand{\Bfin}{\mathcal{B}^{\Delta t}}
\newcommand{\BfinbfU}[1]{\mathbf{B}^{\Delta t}_{\left[ #1 \right]}}
\newcommand{\Bfinbf}{\mathbf{B}^{\Delta t}}
\newcommand{\flow}{\Phi^{\Delta t}}
\newcommand{\R}{\mathbb{R}}
\newcommand{\Rabi}{\Omega_{R}}
\DeclareMathOperator{\meV}{meV}
\DeclareMathOperator{\Id}{\Id}
\newcommand{\D}{\mathcal{D}}
\newcommand{\Dbf}{\mathbf{D}}
\newcommand{\Omegabf}{\mathbf{\Omega}}
\newcommand{\Ubf}{\mathbf{U}}
\DeclareMathOperator{\FWHM}{FWHM}

\renewcommand{\iedlistdecl}{\settowidth{\labelwidth}
{0}}

\title{\LARGE \bf
Accelerating the analysis of optical quantum systems using the Koopman operator
}

\author{Anna Hunstig$^{1}$, Sebastian Peitz$^{1}$, Hendrik Rose$^{2}$ and Torsten Meier$^{3}$
\thanks{$^{1}$Anna Hunstig and Sebastian Peitz are with the Department of Computer Science, Paderborn University, Germany {\tt\small hunstiga@mail.upb.de, sebastian.peitz@upb.de}}%
\thanks{$^{2}$Hendrik Rose is with the Institute for Photonic Quantum Systems (PhoQS), Paderborn University, Germany {\tt\small hendrik.rose@upb.de}}%
\thanks{$^{3}$Torsten Meier is with the Department of Physics and Institute for Photonic Quantum Systems (PhoQS), Paderborn University, Germany {\tt\small torsten.meier@upb.de}}%
}

\begin{document}
\maketitle
\thispagestyle{empty}
\pagestyle{empty}

\begin{abstract}
The prediction of photon echoes is a crucial technique for understanding optical quantum systems. However, it typically requires numerous simulations with varying parameters and input pulses, rendering numerical studies computationally expensive. This article investigates the use of data-driven surrogate models based on the Koopman operator to accelerate this process while maintaining accuracy over many time steps. To this end, we employ a bilinear Koopman model using extended dynamic mode decomposition to simulate the optical Bloch equations for an ensemble of inhomogeneously broadened two-level systems. These systems are well suited to describe the excitation of excitonic resonances in semiconductor nanostructures, such as ensembles of semiconductor quantum dots. We conduct a detailed study to determine the number of system simulations required for the resulting data-driven Koopman model to achieve sufficient accuracy across a wide range of parameter settings. We analyze the L2 error
and the relative error of the photon echo peak and investigate how the control positions relate to stabilization. After proper training, our methods can predict the dynamics of the quantum ensemble accurately and with numerical efficiency.
\end{abstract}

\section{INTRODUCTION}
Linearization of nonlinear dynamical systems is a goal in many research areas, ranging from fluid dynamics \cite{fluid_dynamics} to climate models \cite{climate_models} and robotics \cite{robotics}. The Koopman operator \cite{Koopman1931HamiltonianSA} gives a linear representation of a non-linear system by acting on a space of observables of the system instead of the system state space. Global geometric properties can be derived via its eigendecomposition \cite{spectralproperties1, spectralproperties2}. To date, various computational methods have been established, dynamic mode decomposition (DMD) \cite{Schmid2009, Schmid2010, Koopman-DMD-connection, dmd-theory-appl} and its generalization extended dynamic mode decomposition (EDMD) \cite{Williams2015, Williams2016ExtendingDK} being two of the most popular methods. While the Koopman operator has been well-researched even for particular system classes such as measure-preserving or control systems \cite{Brunton2022ModernKoopman, Otto2021Control, Mauroy2020, OffsetFreeMPC}, its application in quantum physics is fairly new. Some recent investigations include \cite{Klus2022} and \cite{Mezić2023}. Quantum-mechanical descriptions of complex systems are usually computationally challenging and numerically expensive. We examine how a Koopman operator-based model can serve as a surrogate model to conduct parameter studies on quantum systems more efficiently.  \newline
The quantum system we consider in this work is an inhomogeneously broadened ensemble of two-level systems (TLS) that describe, e.g. excitonic resonances in semiconductor quantum dots (QD). These systems are known to emit electromagnetic radiation in the form of photon echoes in four-wave-mixing schemes which involve impulsive excitation with two time-delayed optical pulses that entail non-linear dynamics, see \cite{allen1987optical,meier2006coherent}.
Photon echoes have been proposed as a key component for realizing quantum memory protocols \cite{Damon2011}, and recent studies have demonstrated that the emission time of photon echoes can be controlled optically \cite{Kosarev2020, Grisard2023}.
When an excitonic resonance of an inhomogeneously broadened ensemble of quantum dots, called quantum ensemble (QE), is excited by a short optical laser pulse at $t=0$, the macroscopic polarization of the QE dephases. 
This dephasing can be reversed by exciting the ensemble with a second temporally delayed laser pulse at $t=\tau$. The phase conjugation induced by the second pulse causes rephasing of the microscopic polarization, leading to the buildup of a macroscopic polarization at $t=2\tau$.
This transient macroscopic polarization is the source for the emission of a pulse of electromagnetic radiation which is known as the photon echo.
The contribution of this paper is to demonstrate the advantage of a Koopman-operator approach and data-driven approximation algorithm over conventional methods in efficiently predicting optical quantum experiments accurately over long timescales. 

\section{Preliminaries} \label{sec:preliminaries}
\subsection{Quantum physical model} \label{sec:quantum_optics}
\begin{figure} 
\centering
    \begin{subfigure}{0.45 \textwidth}
    \centering
        \includegraphics[width = 1\textwidth, trim = 0cm 0.6cm 0 5cm]{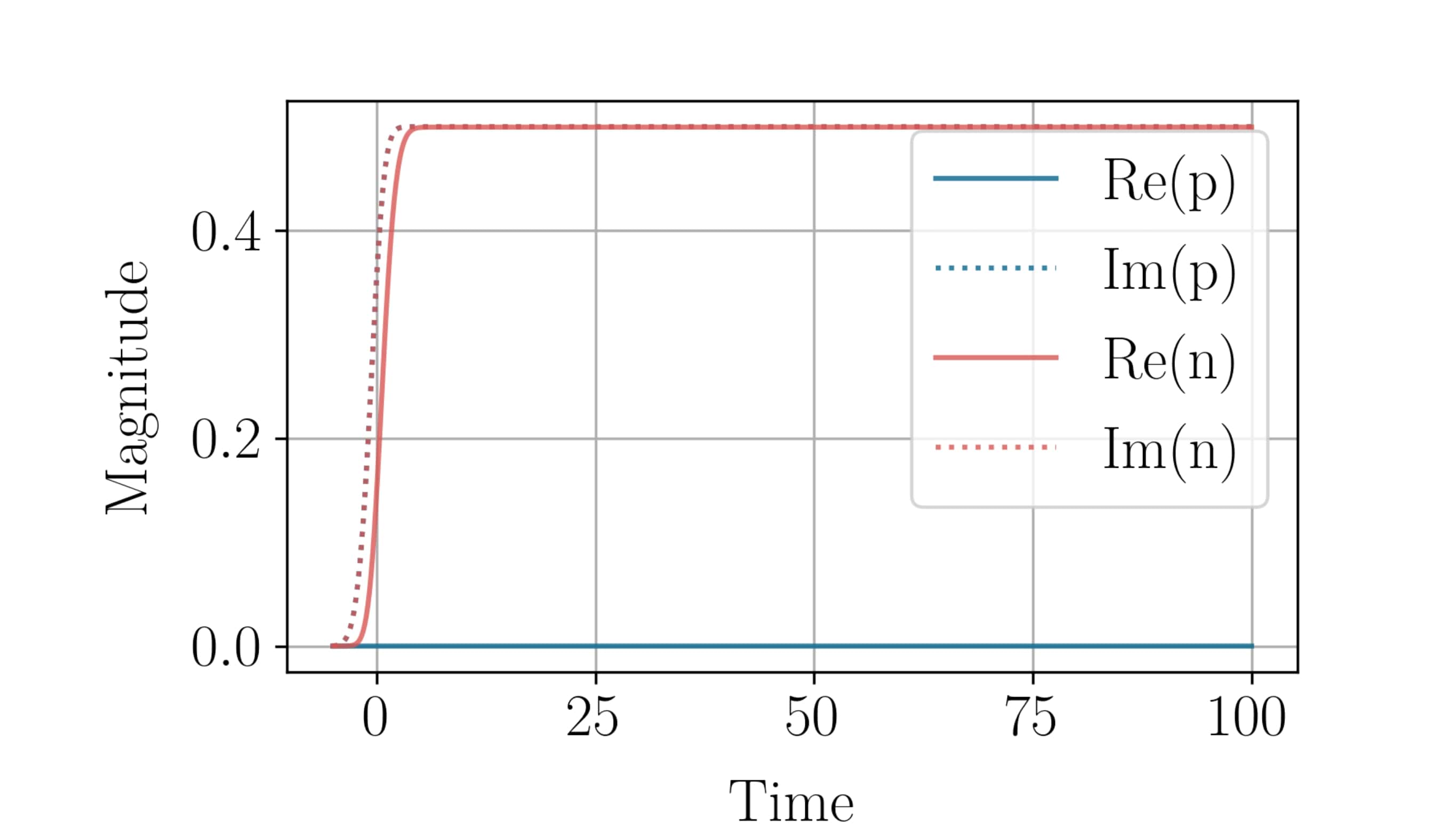}
        \caption{}
    \end{subfigure}
    \begin{subfigure}{0.45 \textwidth}
    \centering
        \includegraphics[width = 1\textwidth, trim = 0cm 0.5cm 0 0cm]{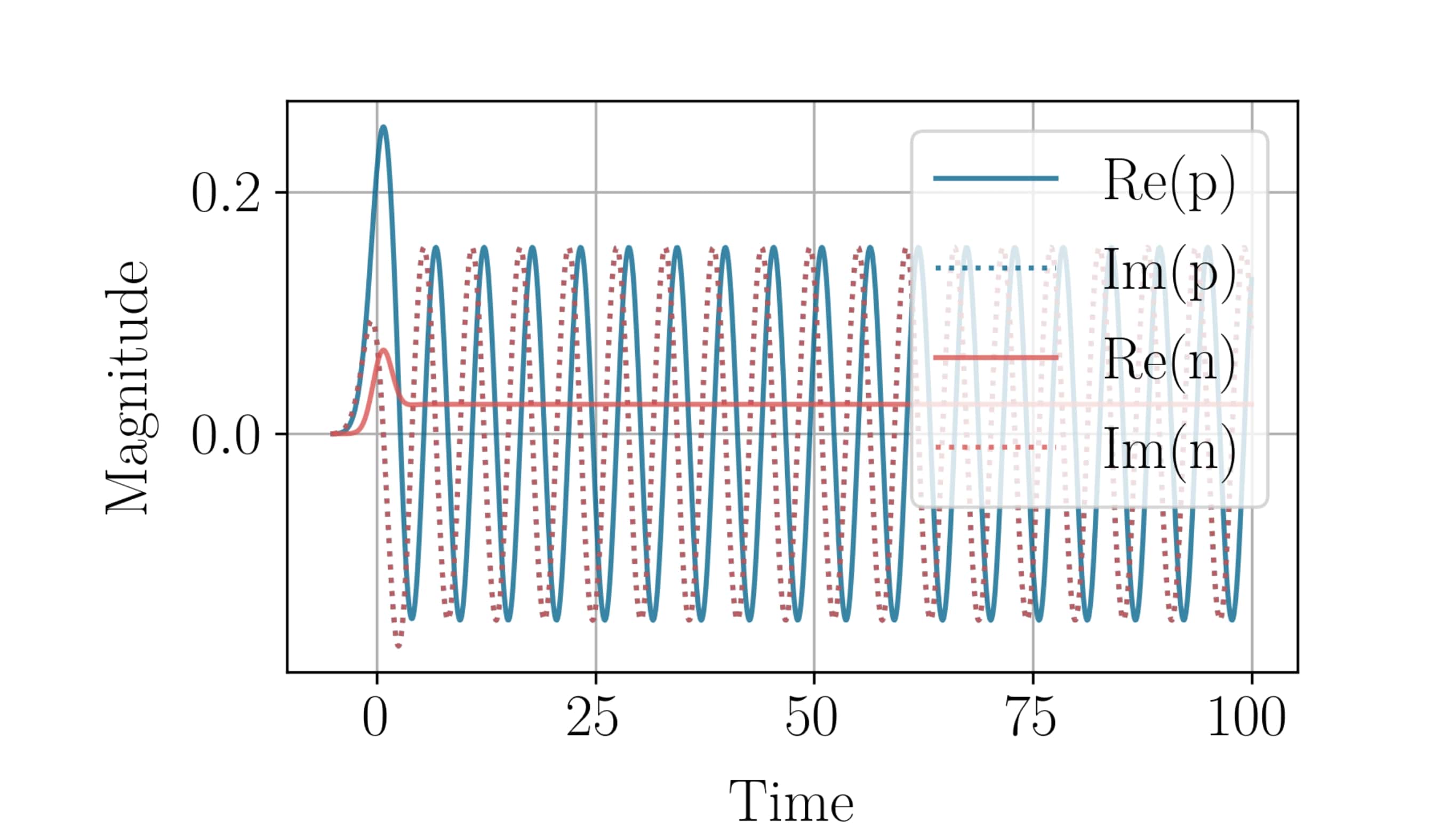}
        \caption{}
        \end{subfigure}
    \caption{Solutions of the OBE with a Gaussian pulse with area $\frac{\pi}{2}$ and duration $2.5$. (a) Detuning $\delta = 0 $ and (b) $\delta = 0.75 \frac{\meV}{\hbar}$.} 
    \label{fig:example_solutions}
\end{figure}

The optical properties and dynamics of near-resonantly excited TLS are obtained by solving the optical Bloch equations (OBE) in the rotating-wave approximation (RWA) \cite{allen1987optical}.
The OBE describe the dynamics of electronic excitations driven by optical fields and can be formulated in terms of the occupation probability $n_{\ell}$ of the energetically higher level and the microscopic polarization or coherence $p_{\ell}$ between the two levels for the $\ell$-th TLS, respectively. In the absence of losses they read:
\begin{equation} \label{eq:OBE}
\begin{bmatrix}
\dot{p}_{\ell}(t) \\ \dot{n}_{\ell}(t)
\end{bmatrix} = 
\begin{bmatrix}
- i \delta_{\ell} p_{\ell}(t) + i \Rabi(t)(1-2 n_{\ell}(t)) \\
2 \Rabi(t) \mathfrak{Im}(p_{\ell}(t))
\end{bmatrix} .
\end{equation}
We denote time in picoseconds and frequency in inverse picoseconds. Besides the two-dimensional complex state ${\xbf_{\ell}(t) = \left(p_{\ell}(t), n_{\ell}(t) \right)},$ the OBE depend on two control quantities. The Rabi frequency ${\Rabi(t)=d E(t)/\hbar}$, proportionate to the electric field $E(t)$ by the dipole matrix element $d$ and the reduced Planck constant $\hbar$, and the optical detuning $\delta_{\ell} =\omega_{\ell} - \omega_L$. Here, $\omega_l$ is the transition frequency of the $\ell$-th TLS and $\omega_L$ the laser frequency. Fig. \ref{fig:example_solutions} shows how variations in $\Rabi(t)$ and $\delta_{\ell}$ affect the dynamics. We are interested in the macroscopic polarization $P(t)$ of the QE, defined as
\begin{equation} \label{eq:def_polarization}
    P(t):= \sum^{\ntestdelta}_{\ell=1} \sigma(\delta_{\ell}) p_{\ell}(t),
\end{equation}
where $\ntestdelta$ represents the number of two-level systems (TLS) in the QE and $\sigma$ denotes the weight distribution of the detunings. This distribution is characterized by its full width at half maximum (FWHM), which corresponds to the inhomogeneous broadening of the detunings. It is important to note that $P(t)$ is dimensionless. The physical polarization can be obtained by multiplying $P(t)$ with $d$. For a discrete set of detunings $\D$ we use the Gaussian distribution $\sigma: \D \rightarrow \R$
\begin{equation} \label{eq:sigma}
    \sigma(\delta_{\ell}) = \exp{-\frac{1}{2} \left( \frac{ 2 \sqrt{2 \ln{2}} \, \hbar \delta_{\ell}}{\FWHM} \right)^2}.
\end{equation} 
The set of detunings is a linearly spaced, discrete array $\D = \left[-R, \dots, R\right] \frac{\meV}{\hbar}$, with $R$ termed the \textit{range}. Proper selection of hyperparameters $R$ and $\FWHM$ ensures accurate modeling of the physical experiment, capturing all relevant dynamics numerically. Choosing $R$ too small compared to $\FWHM$ may result in the domain of $\sigma$ not being fully represented within $\D$. Although a broader range would fully encompass $\sigma$, it could reduce resolution for critical detunings relevant to the pulse durations used in our experiments. The frequency discretization causes unphysical repetitions in the polarization post-photon echo, known as \textit{revival}. The revival time depends on the emission times and frequencies of the laser pulses, and weight distribution. For the pulses used in \cite{Kosarev2020} and simulated in section \ref{sec:simulation1} it can be calculated apriori via 
\begin{equation} \label{eq:revivaltime}
    T_{Rev} = \hbar \frac{\pi (\ntestdelta -1)}{R \cdot 1\mathrm{meV}},
\end{equation} 
which evaluates to $T_{Rev} \approx 110.1$ for $\ntestdelta = 800$ and $R = 15$.

\subsection{The Koopman operator} \label{sec:koopman_operator}
We consider a continuous-time control system 
\begin{equation} \label{eq:DGL}
    \dot{\xbf}(t) = f(\xbf(t),\ubf(t))
\end{equation} with states $\xbf(t) \in \X \subseteq \mathbb{R}^{\dimX},$ control ${\ubf(t) \in \U \subseteq \mathbb{R}^{\dimU}}$ and dynamics $f: \X \times \U \rightarrow \mathbb{R}^{\dimX}$. Let $\ubf(t)=\ubf$ be constant. We can equivalently describe the dynamics by
\begin{equation} \label{eq:flow}
    \xbf(t + \Delta t) = \flow_{\ubf} (\xbf(t)),
\end{equation} where $\flow_{\ubf} : \X \rightarrow \X$ is the flow map with input $\ubf$. Let $\S$ be a space of $L^2$-functions on $\X$ that is closed under composition with the flow map. Then the family $\left(\Kfin_{\ubf} \right)_{\Delta t \geq 0}$ of Koopman operators on $\S$ is defined by
\begin{equation} \label{eq:def_finite_time_koopman_operators}
    \Kfin_{\ubf} \psi := \psi \circ \flow_u, \quad \psi \in \mathcal{S}.
\end{equation} The Koopman family is a semigroup if $\left( \flow_{\ubf} \right)_{\Delta t \geq 0}$ is a semigroup, see \cite{Farkas2020RegularSpaces}.
The generator of the semigroup \cite{KNP+20} is defined by
\begin{equation} \label{eq:def_koopman_generator}
    \K_{\ubf} \psi := \lim_{\Delta t \rightarrow 0} \frac{\Kfin_{\ubf} \psi - \psi}{\Delta t},
\end{equation} for all $\psi \in \S$ such that this limit exists. Inserting \eqref{eq:def_finite_time_koopman_operators} into $\eqref{eq:def_koopman_generator}$ and recalling the definition of the derivative one obtains
\begin{equation} \label{eq:koopman_generator_funkgleichung}
    \K_{\ubf} \psi = f \cdot \nabla_{\xbf} \psi .
\end{equation}
To approximate the Koopman operator, conventionally a set of functions ${\Psi = \{ \psi_1, \dots, \psi_{\dimZ}\}}$ like eigenfunctions or generic basis functions such as polynomials or radial basis functions is chosen to span the function space $\S$.

\subsubsection{Control-affine systems} \label{sec:control_affine_systems}
Suppose the system in \eqref{eq:DGL} is control-affine
\begin{equation} \label{eq:control_affine_dynamics}
    \dot{\xbf}(t) = f(\xbf(t), {\ubf}) = g(\xbf(t)) + \sum_{j=1}^{\dimU} {u}^{j} \, h_j(\xbf(t)),
\end{equation} where $g:\X \rightarrow \X, h_j: \X \rightarrow \X$ may be non-linear and ${u^{j} \in \R}$ denotes the j-th coordinate of the vector $\ubf$. Now, the Koopman ansatz is especially convenient because the control affinity of the system translates to the Koopman generator, resulting in a bilinear model. This will be the main idea for the algorithms in section \ref{sec:algorithms} and was shown in \cite{Peitz2020DataDrivenMP}. To see this, one recognizes that the Koopman generator for the system \eqref{eq:control_affine_dynamics} is given by ${\K_{\ubf} \psi = g \cdot \nabla_{\xbf} \psi + \sum_{j = 1}^p {u}^j h_j \cdot \nabla_{\xbf} \psi}$, compare Eq. $\eqref{eq:koopman_generator_funkgleichung}$. When using a linear combination of $\nlcomb$ constant controls ${\ubf}_1, \dots, {\ubf}_{\nlcomb}$ with ${\ubf = \sum_{i=1}^{\nlcomb} \alpha_i {\ubf}_i}$ and ${\alpha_i \in \R}$ this becomes
\begin{equation} \label{eq:linearcombination_controls}
\begin{aligned}
  \K_{\ubf} \psi &= g \cdot \nabla_{\xbf} \psi + \sum_{j=1}^{\dimU} \sum_{i=1}^{\nlcomb} \alpha_i u_i^j h_j \cdot \nabla_{\xbf} \psi \\
    &= \K_{\zerobf} \psi + \sum_{i=1}^{\nlcomb} \alpha_i \sum_{j=1}^{\dimU}   u_i^j h_j \cdot \nabla_{\xbf} \psi,
\end{aligned}
\end{equation} where we rearranged the sum and used that ${\K_{\zerobf} \psi = g \cdot \nabla_{\xbf} \psi}$ is the case of the unactuated system. By defining ${\B_{\ubf}:= \K_{\ubf} - \K_{\zerobf}}$ we obtain ${\B_{\ubf_i} \psi = \sum_{j=1}^{\dimU}   u_i^j h_j \cdot \nabla_{\xbf} \psi}$ and \eqref{eq:linearcombination_controls} can be expressed as 
\begin{equation} \label{eq:bilinear_model}
    \K_{\ubf} \psi = \left( \K_{\zerobf} + \sum_{i=1}^{\nlcomb} \alpha_i \B_{{\ubf}_i} \right) \psi.
\end{equation}
This system is called \textit{bilinear} in literature, designating the linearity in $\psi$ and affine-linearity in $u$. It holds analogously for time-dependent controls, i.e. the system wrt. control $\ubf(t) = \sum_{i=1}^{\dimU} \alpha_i(t)\ubf_i$ is \begin{equation} \label{eq:bilinear_system_time_dependent}
  \K_{\ubf(t)} \psi(\xbf(t)) = \left( \K_{\zerobf} + \sum_{i=1}^{\dimU} \alpha_i(t) \B_{\ubf_i} \right) \psi(\xbf(t)). \end{equation} \\
Note that, Eq. \eqref{eq:bilinear_model} and \eqref{eq:bilinear_system_time_dependent}  employ the Koopman \textit{generator}. The finite-time operators are not control-affine and introduce a first-order error term:
\begin{equation} \label{eq:finite_time_operator_identity}
    \Kfin_{\ubf(t)} = \Kfin_{\zerobf} + \sum_{i=1}^{\dimU} \alpha_i(t) \Bfin_{\ubf_i} + \O(\Delta t^2).
\end{equation} 
See \cite{Peitz2020DataDrivenMP} for a proof. While this error occurs even if the underlying system is bilinear, as in the case of the OBE, the finite-time Koopman operators bear the advantage that integration in time can be omitted, leading to a likely more efficient computation.

\section{Algorithms} \label{sec:algorithms}
We propose two algorithms based on \cite{Peitz2020DataDrivenMP} and Eq. \eqref{eq:finite_time_operator_identity} to construct a bilinear Koopman model for control-affine systems, referred to as BilinearEDMDc. DMD \cite{Schmid2010} is a popular, computationally efficient method for approximating the Koopman generator using linear state space observations. EDMD \cite{Williams2015} extends this to non-linear observables by using a predefined set of non-linear basis functions, called \emph{dictionary}. Because the OBE are linear in the system state, non-linear observations are not necessary. We present algorithms for general non-linear systems, so we maintain EDMD with a dictionary of monomials up to order 1. Various approaches exist to adapt EDMD for control systems \cite{Proctor2016, Korda2018, Williams2016ExtendingDK}. BilinearEDMDc follows the idea in \cite{Peitz2019} by learning Koopman generators with constant controls and leveraging the control affinity of Koopman operators to construct a bilinear model. Although the OBE \eqref{eq:OBE} form a bilinear system, and hence the Koopman generators could be calculated exactly, we use the finite-time operator approach for easier generalization to nonlinear systems with unspecified dynamics. We first outline BilinearEDMDc to approximate the model \eqref{eq:finite_time_operator_identity} in \ref{sec:bilinear EDMDc} and follow with our algorithms called BE and BERG that are tuned to the OBE. For all algorithms, we denote the set of training control vectors by $\Ubf$ and define $c:= |\Ubf|.$ The total number of distinct control points is $c+1$ as the zero-control $\zerobf$ is always necessary. We assume $\zerobf \notin \Ubf$ to avoid redundancy. The number $\dimZ$ denotes the number of dictionary functions $|\Psi|$, equal to the dimension of the lifted state vector $\Psi(\xbf(t))$. The approximations of the operators $\Kfin_{\ubf}$ are the matrices $\Kfinbf_{\ubf}$.
\begin{figure}[h!]
    \begin{subfigure}{0.5 \textwidth}
    \centering
    \includegraphics[width = 0.9 \textwidth, trim = {1cm 1cm 0 0}, clip]{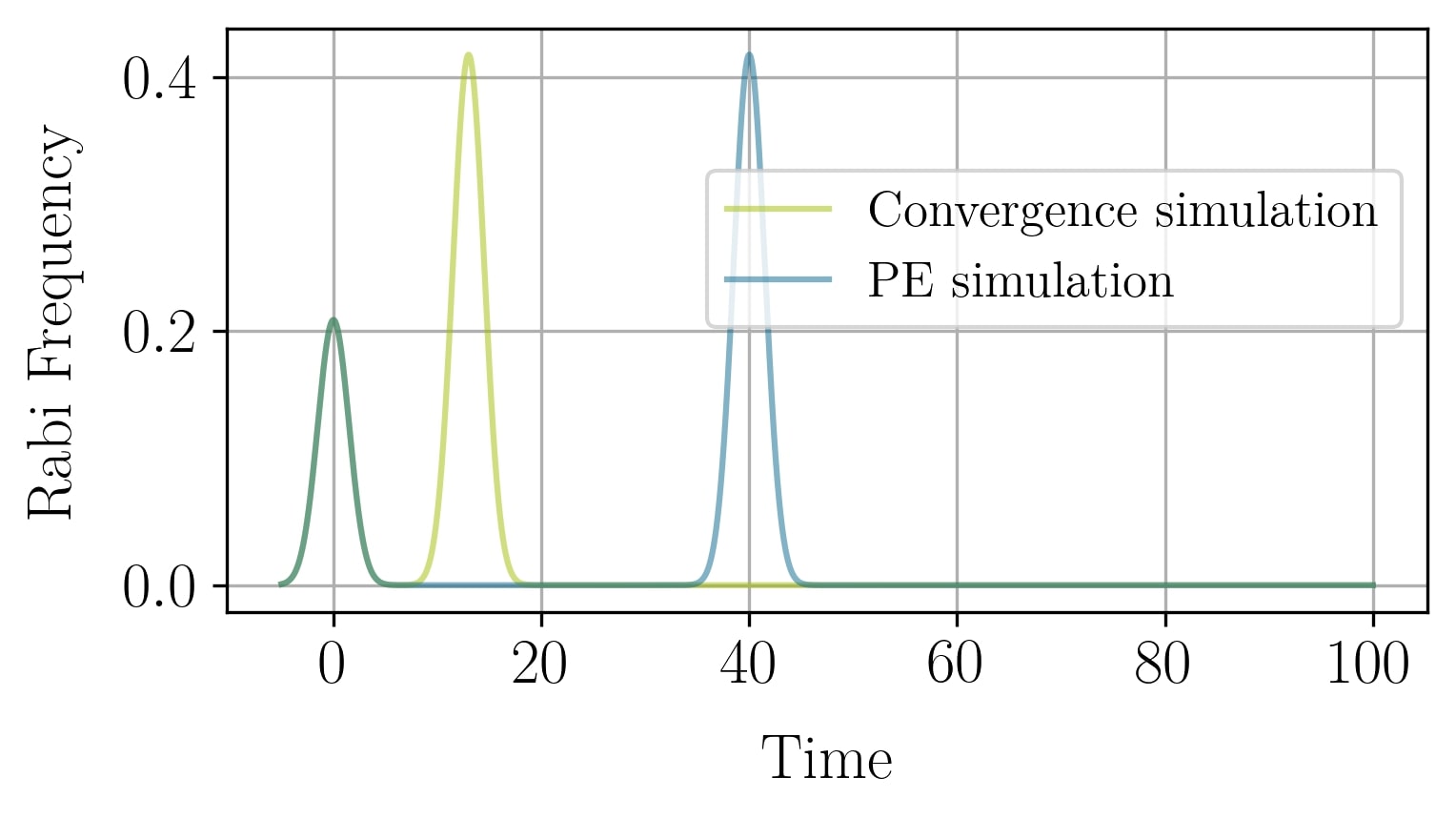}
        \caption{}
        \label{fig:cp_control}
    \end{subfigure}\\
    \begin{subfigure}{0.5 \textwidth}
\centering
       \includegraphics[width = 0.9\textwidth, trim = {1cm 1cm 0 0}, clip]{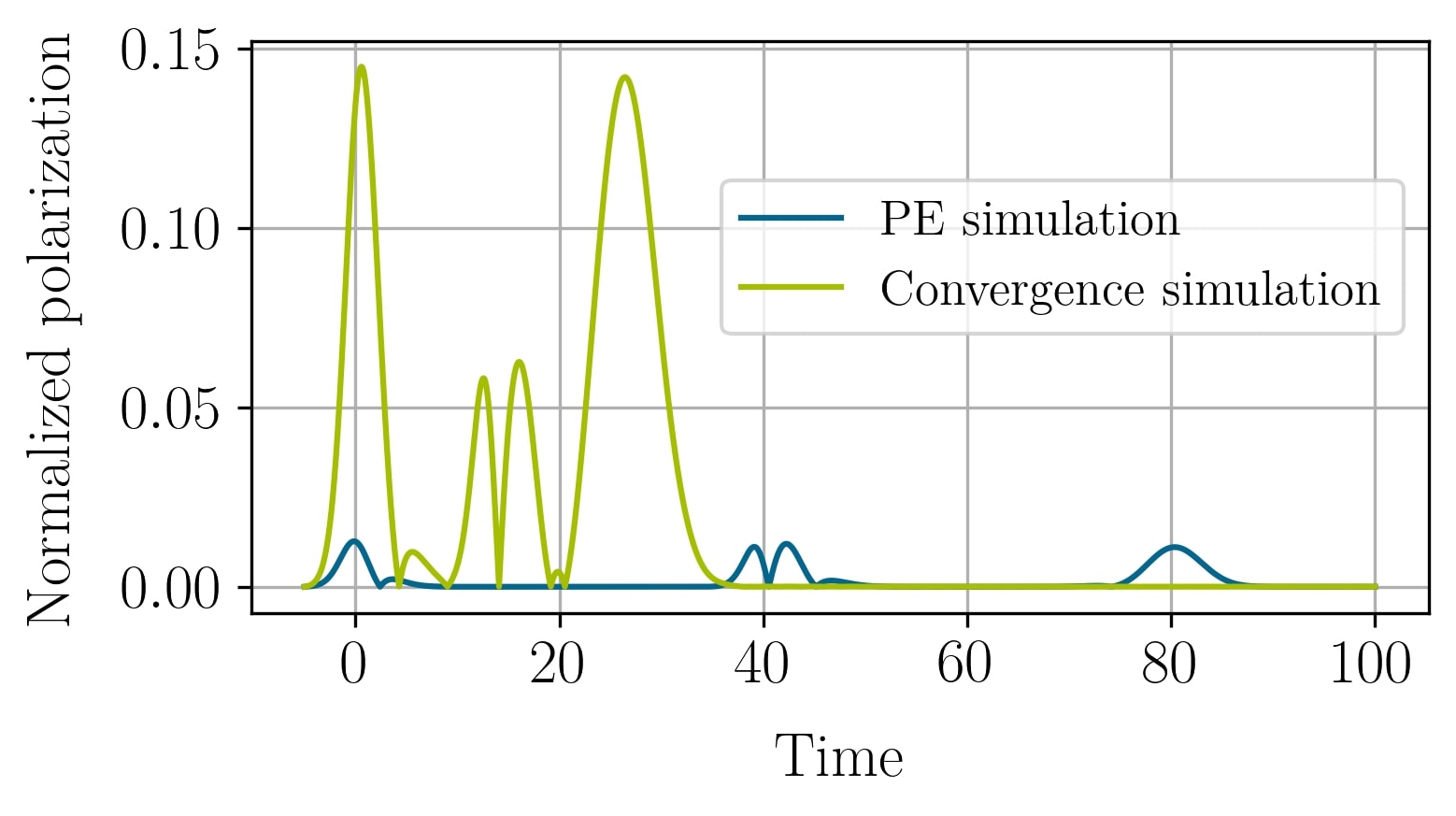}
        \caption{}
        \label{fig:cp_polarization}
    \end{subfigure}
    \caption{(a) Rabi frequency of control pulse and (b) normalized polarization of the simulations in Section \ref{sec:results}.}
    \label{fig:control_polarization}
\end{figure}
\subsection{Bilinear EDMDc} \label{sec:bilinear EDMDc}
The \textit{training} of bilinear EDMDc works as follows: 
\begin{enumerate}
    \item Fix dictionary $\Psi$, control points $\Ubf$ and time step $\Delta t$.
    \item Collect initial system states $\xbf_j := \xbf(t_j)$ (not necessarily consecutive) in a matrix $\Xbf$ 
    and lift $\Xbf$ to 
    $$ \Psi(\Xbf)= \begin{bmatrix}
    \vert & & \vert  \\
    \Psi(\xbf_1) & \dots & \Psi(\xbf_{\ntraintime}) \\
    \vert & & \vert
    \end{bmatrix} \in \R^{\dimZ \times \ntraintime}. $$
    \item Calculate $\mathbf{G}:= \left( \Psi (\Xbf) \Psi(\Xbf)^{\top} \right)^{+},$ where ${+}$ denotes the Pseudoinverse.
    \item For every control $\ubf \in \Ubf$ and $\ubf = \zerobf$:
    \begin{enumerate}
    \item Generate the propagated system states ${\xbf'_{\ubf, j} := \flow_{\ubf} (\xbf_j)}$ and build into matrices $\Xbf'_u$.
    \item Calculate $\mathbf{A_{\ubf}}:= \Psi(\mathbf{X}) \Psi(\Xbf'_{\ubf})^{\top}.$
    \item Approximate the Koopman operator
    $ {\Kfinbf_{\ubf} = \left( \mathbf{G}\mathbf{A_{\ubf}}\right)^{\top} =  \Psi(\Xbf'_{\ubf}) \Psi(\Xbf)^{+}}$.
    \end{enumerate} 
    \item Compute $\mathbf{C} = \Xbf \Psi(\Xbf)^{+}$ to map back to original space.
\end{enumerate} 
\textit{Prediction} entails the following steps:
\begin{enumerate}
    \item Determine the first lifted state $\Psi(\xbf_0).$
    \item For every test timestep $t_k$
    \begin{enumerate}
        \item Represent the test control $\smash{\scalebox{0.95}{$\ubf(t_k) = \sum_{i=1}^{\ncontrols} \alpha_i(t_k) \ubf_i$ }}$ as a linear combination of the trained controls $\ubf_i \in \Ubf$.
        \item Compute the respective Koopman matrix ${\Kfinbf_{\ubf(t_k)} = \Kfinbf_{\mathbf{0}} + \sum_{i=1}^{\ncontrols} \alpha_i(t_k) \Bfinbf_{\ubf_i}}$.
        \item Predict next timestep $\Psi(\xbf_{k+1}) = \Kfinbf_{\ubf(t_k)} (\Psi(\xbf_k)).$
    \end{enumerate}
    \item Project to original state space by $\Xbf = \mathbf{C} \Psi(\Xbf)$.
\end{enumerate}

\subsection{BE and BERG} \label{sec:BERG}
We decompose the two-dimensional complex state space $\xbf(t) = (p(t), n(t))$ into real and imaginary parts, resulting in a four-dimensional real vector. We aim to learn a bilinear model wrt. the control vector $u(t) = \left[\Rabi(t), \delta \right]$. Let $\Omegabf$ and $\Dbf$ be the discretized grids for $\Rabi$ and $\delta$, respectively, with $\ntraindelta := |\Dbf|$. The size of the training set $\Ubf = \Omegabf \times \Dbf$ affects the accuracy and efficiency of the model. BE and BERG both use $\ntraintime = 100$ random initial states $\xbf_j \in \left[-1, 1 \right]^4$ to learn the Koopman matrices and solve the OBE with the same Runge-Kutta method of order four and time step $\Delta t$.

\subsubsection{BilinearEDMDc (BE)}
We use the standard unit vectors $\Ubf = \{\left[1,0\right],\left[0,1\right]\}$, $\ncontrols = 2$ as control points. 
We may write ${\ubf(t_k) = \Rabi(t_k) \left[ 1, 0 \right] + \delta \left[ 0, 1 \right]}$ and see that the prediction at time point $t_k$ is given by
\begin{equation} \label{eq:BE-model}
    \Kfinbf_{\ubf(t_k)} \Psi(\xbf_k) = \left( \KfinbfU{0,0} + \Rabi(t_k)\BfinbfU{1,0} + \delta \BfinbfU{0,1} \right) \Psi(\xbf_k),
\end{equation} compare \eqref{eq:finite_time_operator_identity}. Section \ref{sec:simulation1} shows that this model is not stable for the hyperparameters used in the photon echo experiment, likely due to the inaccuracies introduced by the first-order error from the finite-time operators. Prediction with BE on different parameter settings suggested that it was in particular the rapid fluctuations in the dynamics caused by the wide range $R$ that led to inaccuracies. This motivated the introduction of additional control points.

\begin{figure}[h!]
\begin{subfigure}{0.5 \textwidth}
 \includegraphics[width = 0.9 \textwidth, trim = {0.2cm 0cm 0cm 3cm}, clip]{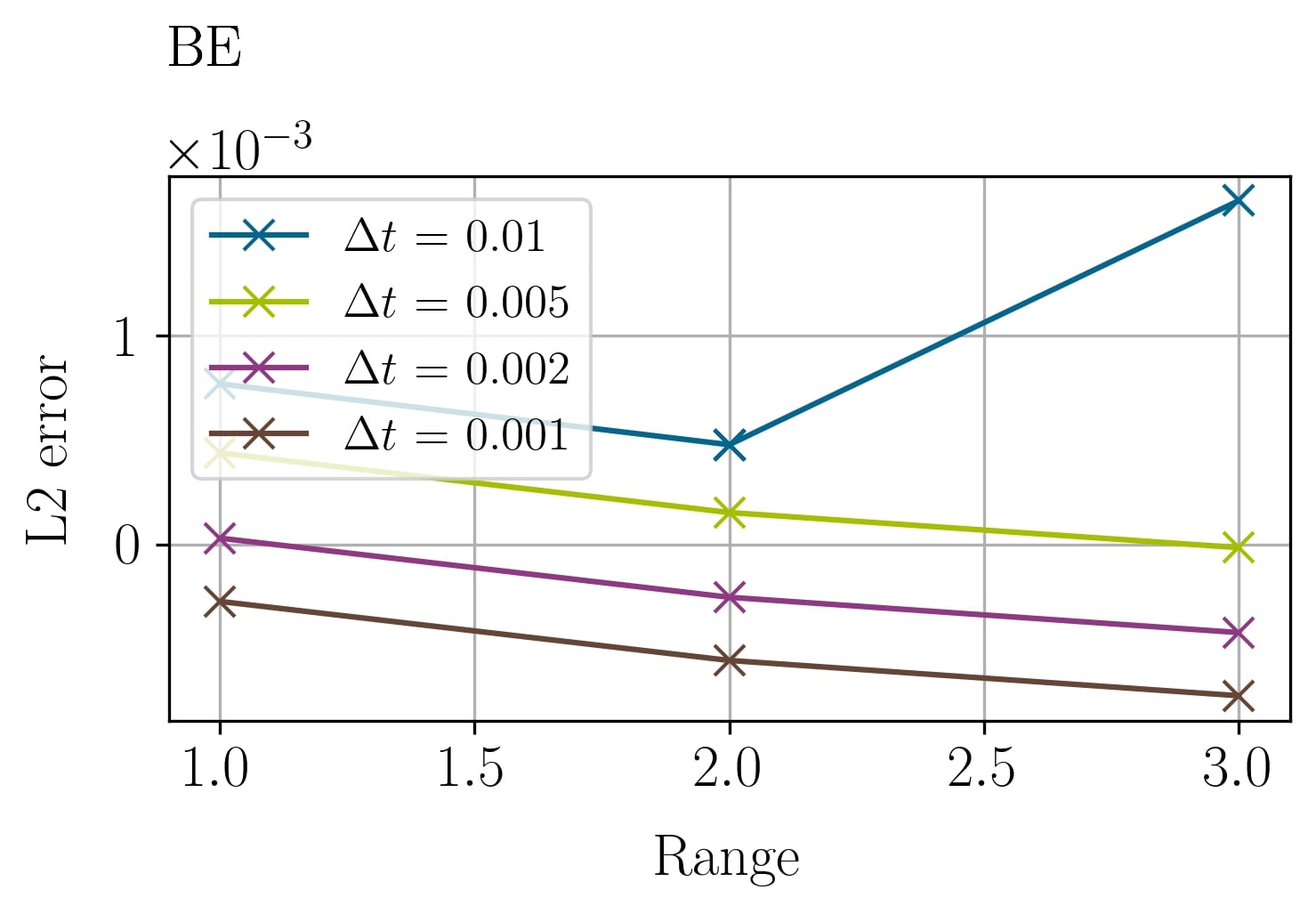}
    \caption{}
    \label{fig:sim1_BE_l2error2}
\end{subfigure}
\begin{subfigure}{0.5 \textwidth}
    \includegraphics[width = 0.9 \textwidth, trim = {0 0 0 0.25cm}, clip]{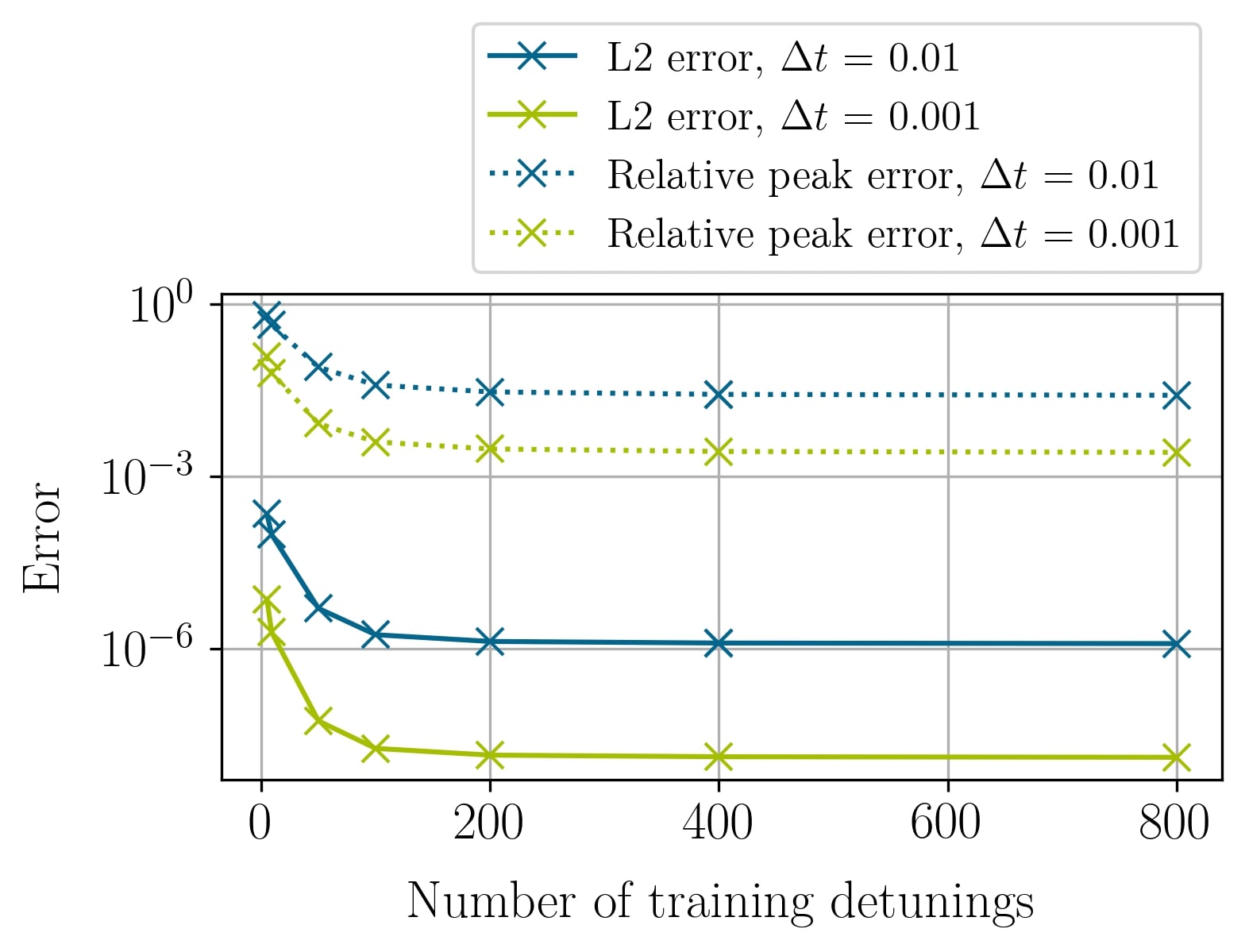}
    \caption{}
\label{fig:sim1_BERG_results}
\end{subfigure}
\caption{(a) L2 error of BE against detuning range and for different step sizes. (b) L2 error and relative peak error of BERG in the photon echo simulation for different step sizes.}
\label{}
\end{figure}

\subsubsection{BE on refined grid (BERG)}
We maintain $\Omegabf = \{0,1\}$, but refine $\Dbf$ to consist of $\ntraindelta$ values between $-R$ and $R$. That is, $\Dbf$ is a coarser instantiation of the test detunings $\D$. In contrast to the single, global BE model \eqref{eq:BE-model} we now construct a family of bilinear models parameterized by $\Dbf$. Consider a fixed test control $u_i = \left[ \omega_i, \delta_i \right]$ and let ${\delta_{1}, \delta_{2} \in \Dbf}$ be the two training detunings that are closest to $\delta_i$ wrt. Euclidean distance. Then $\delta_i = (1-a)\delta_{1} + a\delta_{2}$ for some $a \in \left[ 0,1 \right]$ and ${\left[ \omega_i, \delta_i\right] = \omega_i \left[ 1,0\right] + (1-a) \left[ 0, \delta_{1} \right] + a \left[ 0, \delta_{2}\right]}$. Thereby,
\begin{align*}
\KfinbfU{\omega_i, \delta_i} \Psi(\xbf_k) &= \KfinbfU{0,0} \Psi(\xbf_k)  + \omega_i \BfinbfU{1,0} \Psi(\xbf_k)  \\
&+(1-a) \BfinbfU{0, \delta_1} \Psi(\xbf_k) + a \BfinbfU{0, \delta_2} \Psi(\xbf_k) \\
    &= \left( \KfinbfU{0,0} + \omega_i \BfinbfU{1,0} + \BfinbfU{0, \delta} \right) \Psi(\xbf_k),
\end{align*}
where we abbreviated $\BfinbfU{0, \delta} := (1-a) \BfinbfU{0, \delta_1} + a \BfinbfU{0, \delta_2}.$
\begin{figure}[h!]
        \includegraphics[width = 0.45 \textwidth]{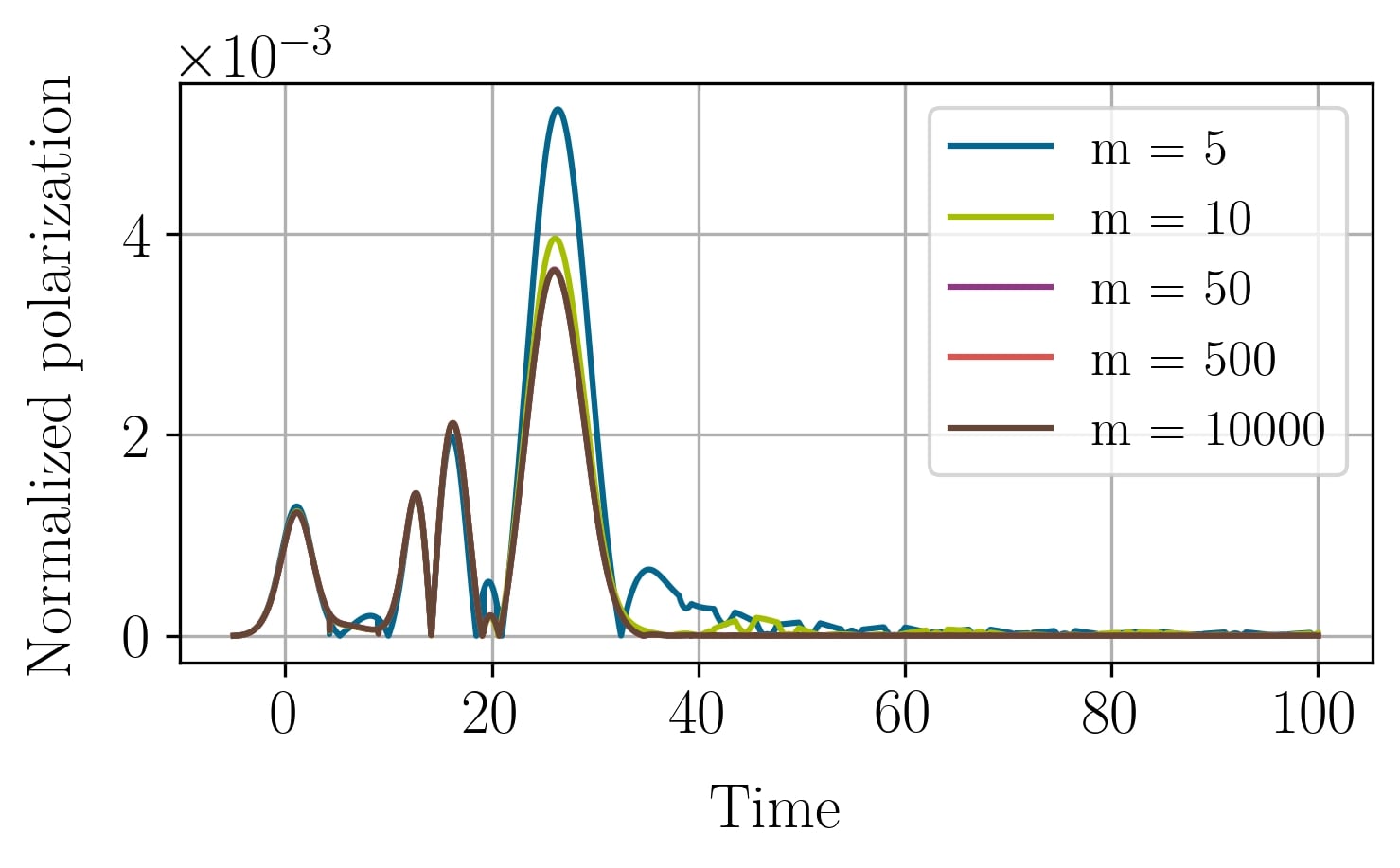}
    \caption{Difference between normalized polarization of reference and BERG in the second simulation.}
   \label{fig:sim2_BERG_pol_difference}
\end{figure}

\begin{figure*}
    \centerline{
    \begin{subfigure}{0.45 \textwidth}
        \includegraphics[width = 1 \textwidth, trim = {0 0 0 0cm}, clip]{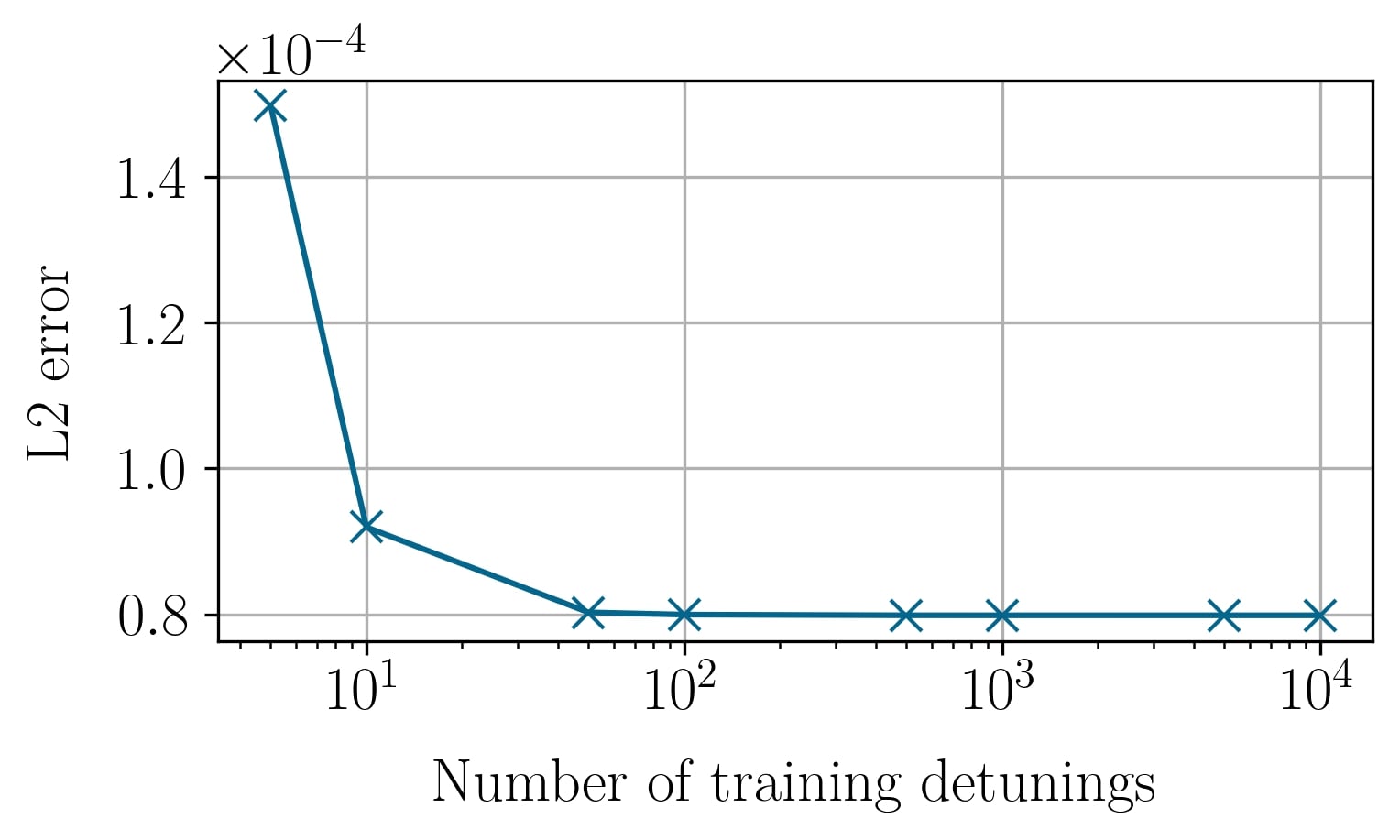}
        \caption{}
        \label{fig:sim2_BERG_l2error}
    \end{subfigure}
    \hfil
    \begin{subfigure}{0.45 \textwidth}
        \includegraphics[width = 1 \textwidth, trim = {0 0 0 0cm}, clip]{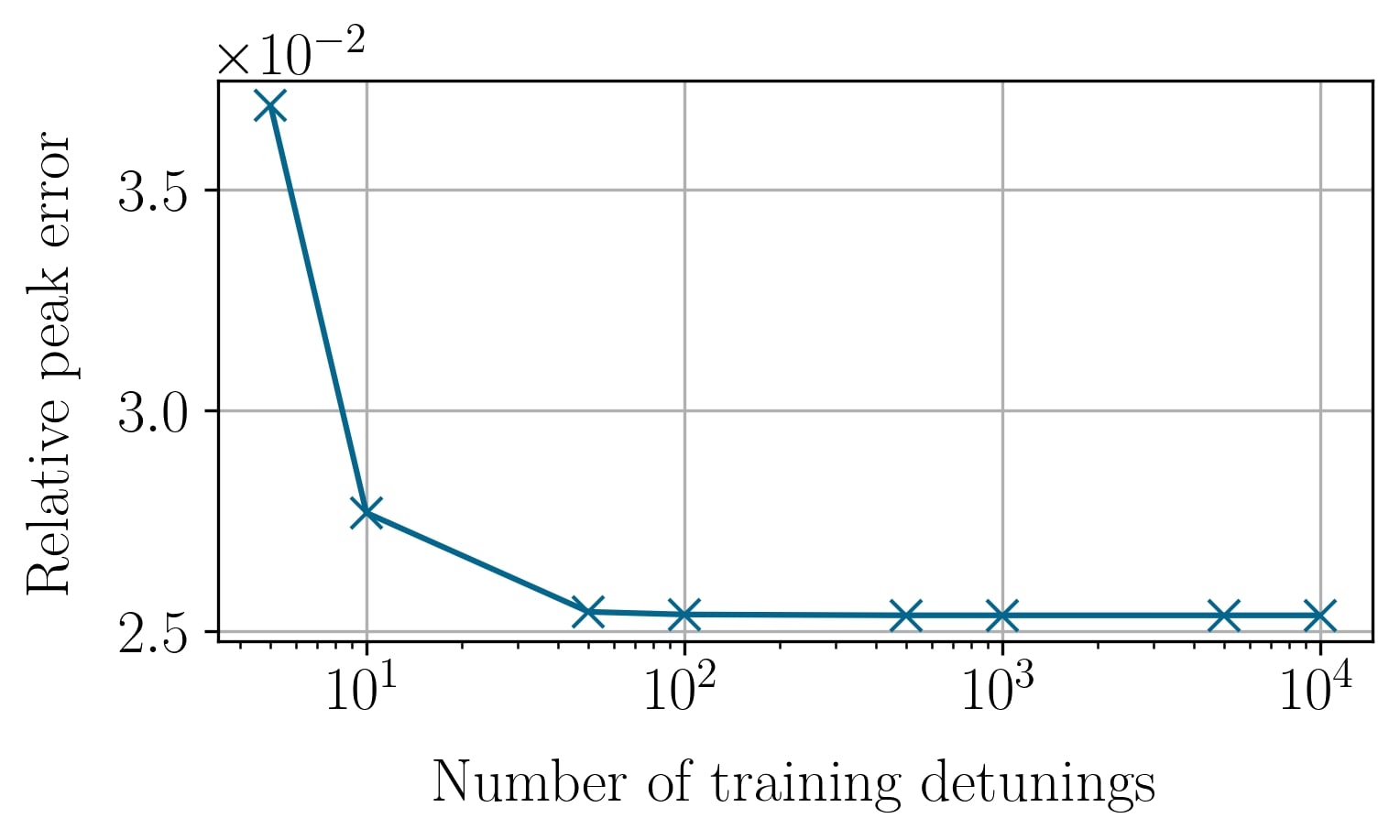}
        \caption{}
        \label{fig:sim2_BERG_relpeakerror}
    \end{subfigure}}
    \caption{(a) L2 error and (b) relative peak error for BERG in the convergence simulation. In both cases, $\ntraindelta + 1$ detunings were trained, where $\ntraindelta \in \{ 5, 10, 50, 100, 500, 1000, 5000, 10000 \}$.}
    \label{fig:BERG_sim2_results1}
\end{figure*}
\section{Results} \label{sec:results}
\subsection{The photon echo} \label{sec:simulation1} 
Our ultimate goal is to learn a model that accurately simulates the dynamics of the OBE in the photon echo experiment \cite{Kosarev2020}. In this setting, the QE consists of $\ntestdelta = 800$ TLS with evenly spaced detunings in the range $R=15$ and a weight distribution $\sigma$ defined by $\FWHM = 7.5$~meV. The exciting signal is composed of two pulses, each with duration $2.5$. The initial pulse has pulse area $\frac{\pi}{2}$ and is centered at $t = 0$, the second is a $\pi$-pulse, centered at $t = 40$, see Fig.~\ref{fig:cp_control}. The reference data is computed with the initial value solver from SciPy \cite{scipy}, using RK45 and parameters $\Delta t = 0.01$, $\text{rtol} = 1e-8$ and $\text{atol} = 1e-11$ for the interpolation. See  Fig.~\ref{fig:cp_polarization} for a visualization of the reference polarization \eqref{eq:def_polarization}. Evaluation occurs from $t=-5$ to $t=100$ and we select two error measures that remain invariant under the size of the ensemble $\ntestdelta$, as this variable will vary in the second simulation, Section \ref{sec:simulation2}. We aim to quantify both the overall error and the deviation of the model in the photon echo peak. Let $\Bar{P}(t) = \frac{P(t)}{\ntestdelta}$ be the normalized polarization, then the respective L2 error is given by
\begin{equation*}
    L_2(\Bar{P}_{Ref}, \Bar{P}_{BE}) = \int^{t=100}_{t=-5} \left( \frac{P_{Ref}(t)}{\ntestdelta} - \frac{P_{BE}(t)}{\ntestdelta} \right)^2 dt,
\end{equation*} 
where $P_{Ref}$ and $P_{BE}$ denote the macroscopic polarization of the reference and BE, respectively. We write $S_{Ref}$ and $S_{BE}$ for the photon echo peak of the reference and BE and denote the relative peak error
\begin{equation*}
    L_{peak}(P_{Ref}, P_{BE}) = \left| \frac{S_{Ref} - S_{BE}}{S_{Ref}}\right| .
\end{equation*} 
When simulating the photon echo experiment with BE and a step size of $\Delta t = 0.01$, the model deviates drastically from the reference calculation from $t = 2$ onwards.
Although the reduction of the step size to $\Delta t = 0.001$ slightly decreases the L2 error, BE does not suffice to model the original experiment. We observed that the L2 error of the model heavily depends on the detuning range that is used. Varying $R$ showed that the lowest L2 error for BE with $\Delta t = 0.01$ is $9 \times 10^{-5}$ and attained at $R = 2$, see Fig. \ref{fig:sim1_BE_l2error2}. With these parameters BE matches the reference trajectory and approaches the correct photon echo peak, which is supported by a corresponding relative peak error of $0.05$. However, for greater ranges $R \geq 4$, the L2 error diverges. BE is therefore not relevant for the prediction of the OBE in general. We find that, although the underlying dynamics are comparatively simple, the large number of time steps, in combination with the control input, poses substantial challenges. This implies that the modeling process needs to be performed judiciously. \\
In BERG, the refinement of training detunings $\Dbf$, i.e. increasing $\ntraindelta$, decreases the error of the operators. The trade-off between $\ntraindelta$ and the error measures is illustrated in Fig. \ref{fig:sim1_BERG_results}. The trajectories for the two error measures exhibit similarity, leading to two conclusions. Firstly, it is evident that the step size significantly influences the errors. Secondly, ${\ntraindelta = 100}$ and ${\ntraindelta = 200}$ provide an excellent trade-off between accuracy and data requirements. For $\Delta t = 0.01$ they lead relative peak errors of approximately $0.039$ and $0.029$, respectively, and L2 errors of $1.75 \times 10^{-6}$ and $1.34 \times 10^{-6}$. By choosing an appropriate number of training detunings, BERG can simulate the photon echo experiment with an error that may be tuned further with step size.
\subsection{Convergence with QE size} \label{sec:simulation2}
The second simulation investigates the convergence of the photon echo with the number of QD in the QE. 
This number $\ntestdelta$ is in reality very large, which incurs high costs for conducting detailed simulations for all QDs. By learning a Koopman model from a subset of trajectories $\ntraindelta \ll \ntestdelta$ trajectories, we can yield a great advantage in efficiency compared to the conventional expensive and time-intensive model.
We fix $\ntestdelta = 10^4$ and $\Delta t = 0.01$ and compare the relative peak error of the Koopman model trained on $\ntraindelta \ll \ntestdelta$ detunings, evaluated on $\ntestdelta$ detunings with the relative peak error of a reference calculation of $\ntestdelta$ detunings. Since the revival time approaches zero for decreasing number $\ntraindelta$ of QD in the simulated QE, i.e. $T_{Rev} \rightarrow 0$ for $ \left( \ntraindelta -1 \right) \rightarrow 0$, the non-physical revival distorts the training data. By tuning the parameters such that $T_{Rev}$ exceeds our evaluation time $t = 100$ even for small $\ntraindelta$, the problem can be circumvented. The resulting parameters are $R = 1$, $\FWHM = 1$~meV, and a pulse as shown in Fig. $\ref{fig:cp_control}$. Fig.~\ref{fig:sim2_BERG_pol_difference} shows the difference between the polarization trajectory calculated by BERG for different $\ntraindelta$ and the reference trajectory. An error visualization can be found in Fig.~\ref{fig:BERG_sim2_results1}. We examine that the trajectories of the L2 error and the relative peak error of BERG are similar to the error data in the photon echo experiment. While the number of trained detunings $\ntraindelta$ now ranges up to $\ntestdelta = 10^4$, we observe a similar, significant dip in the errors. The optimal number of training detunings are $\ntraindelta = 10$ and $\ntraindelta = 50$ with a relative peak error of approx. $0.028$ and $0.025$, respectively, and L2 error of $9.19 \times 10^{-5}$ and $8.03 \times 10^{-5}$, implying that we have reduced the number of expensive simulations by a factor of $1000$ with BERG.

\addtolength{\textheight}{-1.5cm}   

\section{Conclusions}
We have investigated the possibility of using Koopman operator-based surrogate models to accelerate the analysis of optical quantum systems. Even though the original system has bilinear dynamics, the small inaccuracies of the discrete-time bilinear Koopman model pose significant challenges for long-term predictions such that special care has to be taken during modeling. By introducing a refined training grid for the detuning, the model can be stabilized, leading to accurate predictions of TLS in an RWA. Several aspects of training the bilinear model remain to be explored. For instance, the impact of using alternative sampling techniques for detuning control instances, as opposed to the linearly spaced detunings applied in our approach, has yet to be determined. Additionally, using a more refined grid for the optical pulse could further reduce errors. The growing interest in quantum systems underscores the need for further research into efficient simulation and control methods.
We demonstrated the potential of a Koopman operator-based model for quantum applications by presenting accuracy results on a simple quantum system. However, further investigation is needed to explore the applicability of this model to more complex quantum systems. Examples include the nonlinear optical dynamics of electronic many-body systems or the ultrafast dynamics of matter driven with extremely strong fields, providing access to attosecond time scales.

\section{ACKNOWLEDGMENTS}
AH and SP acknowledge support from the German Federal Ministry of Education and Research (BMBF) within the AI junior research group ``Multicriteria Machine Learning''.
TM and HR acknowledge funding from the Ministry of Culture and Science of the State of North Rhine-Westphalia via the PhoQC project.



\bibliographystyle{IEEE/IEEEtran}
\bibliography{IEEE/IEEEabrv, bibliography}

\end{document}